# Non-intrusive load decomposition based on CNN-LSTM hybrid deep learning model

Xinxin Zhou [1], Jingru Feng [1], Yang Li [2,*]

[1] School of Computer Science, Northeast Electric Power University, Jilin 132012, China

[2] School of Electrical Engineering, Northeast Electric Power University, Jilin 132012, China

* Corresponding author. E-mail address: liyang@neepu.edu.cn (Y. Li).

**ABSTRACT:** With the rapid development of science and technology, the problem of energy load monitoring and decomposition of electrical equipment has been receiving widespread attention from academia and industry. For the purpose of improving the performance of non-intrusive load decomposition, a non-intrusive load decomposition method based on a hybrid deep learning model is proposed. In this method, first of all, the data set is normalized and preprocessed. Secondly, a hybrid deep learning model integrating convolutional neural network (CNN) with long short-term memory network (LSTM) is constructed to fully excavate the spatial and temporal characteristics of load data. Finally, different evaluation indicators are used to analyze the mixture. The model is fully evaluated, and contrasted with the traditional single deep learning model. Experimental results on the open dataset UK-DALE show that the proposed algorithm improves the performance of the whole network system. In this paper, the proposed decomposition method is compared with the existing traditional deep learning load decomposition method. At the same time, compared with the obtained methods: spectral decomposition, EMS, LSTM-RNN, and other algorithms, the accuracy of load decomposition is significantly improved, and the test accuracy reaches 98%.

## 1. Introduction

As the goal of the future development of global power grids, smart grids have developed rapidly whether they are smart grids for large-scale systems or smart home devices for micro-systems. Electricity, as the outstanding energy source, is the basic energy source for development, so reducing the consumption of electric energy significantly. The key to the development of an intelligent electricity strategy is to understand the energy consumption distribution and usage habits of electrical equipment of electricity users. To do this, the load must be flexible and the devices the home turns on at a given time should be determined in real-time, a process known as load monitoring [1]. In 1992, Hart proposed non-intrusive load decomposition (NILM) technology [1], this is an economical and convenient method of monitoring. Different from load decomposition whose main problem is to refine characteristics [2], the core issue of the non-intrusive load decomposition is the selection of unique characteristics to distinguish different appliances, which is the key to reduce the operational cost of building or residential microgrids. Over the past three decades, numerous researches have been devoted to the development of NILM methods that can decompose end-user electricity consumption data measured by smart meters into their appliance level component processes [3]. The NILM technology decomposes the electricity consumption data collected by the user's electricity meter to display the use of household appliances in real time [4], which offers technical assistance for the development of a smart grid power consumption strategy.

Up to now, there have been some pioneering works that attempt to solve the NILM issues. For example, in reference [5], a system for making devices based on classification labels and the total power measurement is proposed. This system can automatically build models for devices, and the error range of results is relatively low. Reference [6] uses an artificial bee colony algorithm to estimate a single power load, which enhances the accuracy of load monitoring and is easy to implement in the existing smart meter technology without modifying the hardware. In reference [7], a load identification method combining wavelet transform and decision tree is introduced, but this method lacks general applicability in the case of multiple loads. Reference [8] puts forward a method based on V-I trajectory. Different loads are distinguished by the shape and color of V-I trajectory, and knowledge between different fields is combined to enhance the uniqueness of load characteristics.

As an emerging machine learning techniques, deep learning has been successfully applied to sovle various engineering problems, such as line defect recognition [9], power quality disturbance classification [10-11], power system short-term voltage stability assessment [12] and image classfication [13,14] and wind power prediction [15-16]. Compared with the V-I image features widely studied in reference [17], the adaptive weighted recursive graph block is proposed to represent the electrical features in the NILM, which improved the generalization performance and guaranteed the uniqueness of appliance features. LSTM networks are widely used in time series analysis because of their ability to handle long-term dependencies [12]. For NILM problems, LSTM networks have a comparative advantage in detecting load data changes, which have been demonstrated in previous related studies [18-20]. Due to the powerful feature learning ability, convolutional neural network (CNN) has also been used in the NILM field. For example, reference [21] adopts the successful deep CNN and sets up a data enhancement technology to solve the second type of household appliances problem of NILM. References [22]-[24] propose more advanced deep learning modes to realize NILM, and these modes constitute the most advanced level in this research field. Compared with load decomposition research, reference [25] provides an electrical application on/off state analysis through deep learning, and has a good transfer learning effect. In addition to deep learning-based NILM methods, many researchers have focused on implementing NILM through shallow machine learning methods. Support vector machine (SVM) [26], decision tree (DT) [1], Moth-flame optimization algorithm [27], Random Forest (RF) [28] and decision bagging tree classifier [29] have been used to solve event-based NILM. The hidden Markov model in reference [30] is used to change the problem into a coding and decoding problem, which can retain the dependency between loads and decompose multi-state loads, and the online load decomposition is realized, but the decomposition accuracy is low. Reference [31] improves the Hidden Markov Model and proposes a solution method of piecewise quadratic constrained programming, which can efficiently solve the model and cooperate with existing smart meters. Based on the framework of the additive factorial hidden Markov model, a combined algorithm of active and reactive power is proposed in reference [32], and the superiority of the result is verified by comparison with the algorithm proposed in reference [1]. In recent years, Wittmann et al. proposes the mixed-integer column programming method for load decomposition and achieves high accuracy [33]. The author of [34] proposed a completely unsupervised NILM method using graph signal processing, which can solve the training overhead and related complexity of traditional methods. The work of NILM based on graph signal processing was further expanded in [35], which solved the problem of measurement noise and the influence of unknown load on the performance of load decomposition and made great progress.

Although there are many non-intrusive load decomposition methods, there are still many problems: it is difficult to model electrical loads with large fluctuations, and there are no more performance analysis indicators used in the evaluation model, or measurement errors occur, and accuracy is not a high-level issue.

Among the above-mentioned research methods, some methods such as Wittmann et al. proposed the mixed-integer sequence planning method without considering the time correlation characteristics of electrical signals. The signals at each moment are decomposed in isolation, and the operating status of each electrical appliance is processed separately. Making full use of time information, this article introduces long short-term memory (LSTM) network into the model. In addition, inspired by reference [36], the convolutional neural network (CNN) is introduced in order to automatically extract load features and not lose lots of information in the process of dimensionality reduction.

A reasonable construction model can enhance the accuracy of the decomposition. For the purpose of better enhance the effectiveness of the deep learning method in the field of non-intrusive load decomposition, the CNN-LSTM hybrid model is introduced into the load decomposition algorithm to realize the extraction of the spatial characteristics and temporal features of electrical power.

The main contributions are highlighted as follows:

1) This paper proposes for the first time an evaluation framework combining CNN-LSTM hybrid model, making full use of the spatial features of convolutional neural network and the temporal features of long-term memory network, taking into account the characteristics of data feature extraction and based on time series. The network structure is simple but powerful, and the recognition error under-sampling is reduced.

2) The test results show that the proposed method is superior to CNN and LSTM and other independent deep learning algorithms, as well as to existing methods in the paper: EMS, LSTM-RNN, and other algorithms, and can improve the decomposition accuracy. Moreover, this hybrid model still maintains high performance under different SNR.

3) In this paper, a series of evaluation indicators such as ACC, MCC, and F1-SCORE were added, and noise impact analysis was also added to better verify the performance of the proposed model.

The main structure of this paper: Section 1 mainly introduces the related background and related research of non-intrusive load decomposition. Section 2 introduces related theories about CNN and LSTM network. Section 3 mainly proposes a CNN-LSTM hybrid model based on non-intrusive load decomposition and introduces its related structure. Section 4 uses UK-DALE data to train and test the proposed hybrid model, and analyze the test results. Section 5 draws the foremost conclusions of this task.

## 2. Problem statement

What is non-intrusive load decomposition, is to be able to achieve the decomposition and analysis of the whole substation, building internal load cluster, further obtain useful information method. One use of non-intrusive load decomposition is for home users to know when each household appliance is on and off and, if possible, how much power each appliance consumes. If you use an example to explain, it is very similar to our common cocktail party problem. What about cocktails? For example, as shown in the picture below, we can hear sounds coming from all directions while walking on the road, and we need to isolate each sound, such as the sound of cars, the sound of people whispering to each other, the sound of animals.

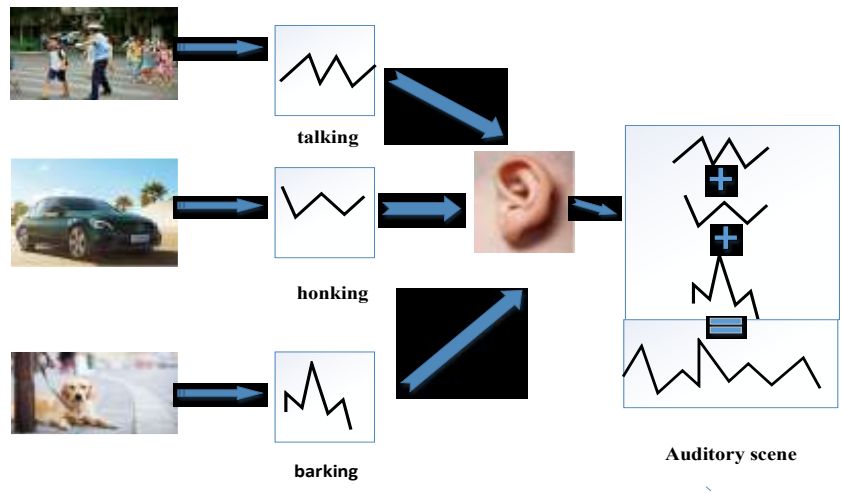

Fig.1. Cocktail problem diagram

Our non-intrusive load decomposition problem is to know how much power each household appliance, such as refrigerator, TV set, and washing machine, consumes from the total electricity meter of the family. With the gradual development of non-intrusive load decomposition technology, its core is to use the least equipment to realize the monitoring of all the electricity load in the residential, so as to provide more detailed household electricity information for the power system.

## 3. Deep learning neural network-related theories

### 3.1 Neural Network

The Neural network is formed of neurons and connections between nodes (synapses). It also connects multiple single neurons and uses the output of one neuron as the input of the next neuron. In real life, it is often a multi-layer network composed of multiple perceptrons. The structure diagram shown in Fig.2 is as follows, this is the classic neural network model.

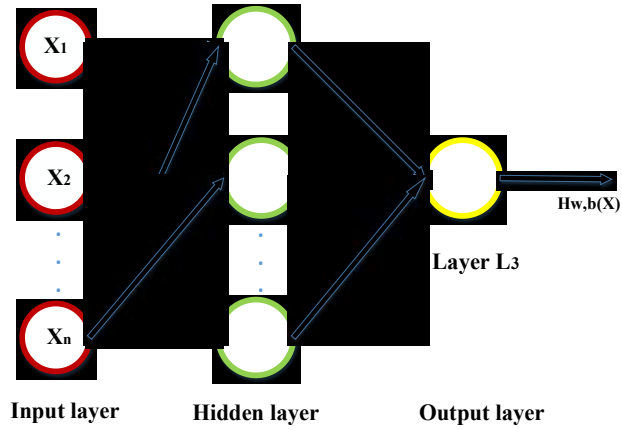

Fig. 2. Neural network structure diagram

Each layer of the network has multiple neural units. The neurons in the upper layer are mapped to the neurons in the next layer through the activation function. There are corresponding weights between them, and the output is our classification category.

### 3.2. CNN

In the 1960s, the biological research of Hubel et al. showed that the transmission of visual information from the retina to the brain was completed by multiple levels of receptive fields (Receptive Field) excitation, and then proposed a CNN. The CNN is a kind of feedforward neural network with a deep structure [37], which is an end-to-end model in Fig. 3.

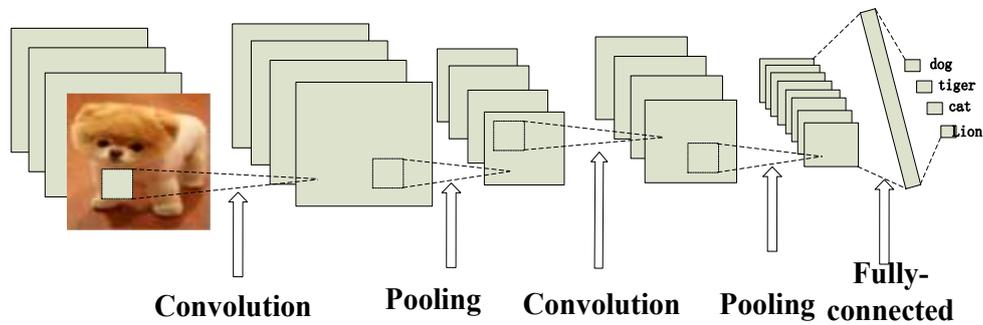

Fig. 3. Schematic diagram of CNN

The calculation formula of the CNN is:

$$N = (W - F + 2P)/S + 1 \quad (1)$$

Among them $N$: the size of the output, $W$: the size of the input, $F$: the size of the convolution kernel, $P$: the size of the padding value, $S$: the size of the step size.

The CNN model is mainly formed of the input layer, convolution layer, down-sampling layer, fully connected layer, and output layer [38].

The work of the convolutional layer is to extract features from the input data [39], which includes the convolution kernel, the convolutional layer parameters, and the activation function. The convolution kernel is an essential part of the parameters of the convolution layer, and the essence of the convolution kernel is the feature extractor.

The activation function is usually used to add nonlinear factors after the convolution kernel [40] because the expressive power of the linear model is not enough. Commonly used activation functions are sigmoid, tanh, relu, etc. The representation of the activation function is as follows:

$$A_{i,j,k}^l = f(Z_{i,j,k}^l) \tag{2}$$

The output value of the convolutional layer is calculated as the following formula:

$$c_{i,j} = F(\sum_{a=0}^{2}\sum_{b=0}^{2} W_{a,b} X_{i+a,j+b} + W_m) \tag{3}$$

Among them, $W_{a,b}$ stands for the weight of row *m* and column *n*, $W_m$ is the offset item of filter, $F(\cdot)$ denotes the activation function, and $X_{i,j}$ is the output vector of the previous layer.

In CNN, the down-sampling layer is used to decrease the spatial dimension but does not reduce the depth of the network. It can keep the most important information and is usually behind the convolutional layer. The feature vectors output by the convolutional layer can be reduced through pooling and the results can be improved at the same time. The two most commonly used pooling methods are average pooling and maximum pooling [41].

In the whole CNN, the fully connected layer is a classifier. It is located at the end of the network and performs regression classification on the features extracted. Therefore, the CNN is mainly composed of two parts, one part is feature extraction (convolution, activation function, pooling), and the other part is classification and recognition (fully connected layer) [42].

### 3.3. LSTM

The recurrent neural network (RNN) has the functions of data learning, classification, prediction, and so on. At the same time, it has a time-series characteristic, which has high efficiency in the prediction of time series. It can only evolve according to the passage of time, so as to predict the data. However, it is difficult for RNN to remember input information that is too far apart, so the long-term dependency problem is the fatal injury of traditional RNN [43].

As a variant of the RNN, the LSTM mainly solves the problem of gradient disappearance, which can make the network remember the content for a longer time and make the network more reliable [12, 44]. LSTM has the capability to strike out or increase information to the cell state. This capability is given by a structure called a gate. LSTM possesses three gates, are input gate, forget gate, and output gate [45], which are used to provide read, write and reset functions respectively. The structure diagram of LSTM in Fig. 4.

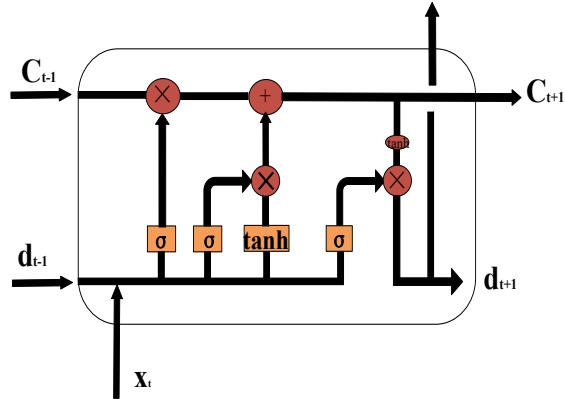

Fig. 4. Schematic diagram of LSTM

Among them, $C_{t-1}$ is the cell state from the previous module, $d_{t-1}$ is the output of the previous module, $X_t$ is the current input, used to generate new memory, and the output information includes the cell state $C_t$ transmitted later, new output $d_t$.

The forgetting gate in LSTM is actually a valve. When the input gate is always open, a lot of information will flood into the memory. At this time, a forgetting mechanism needs to be added to remove the information in the memory. This is the forgetting gate. It looks at $d_{t-1}$ (previous output) and $X_t$ (current input) and outputs a number among 0 with 1 for every digit in the cell state $C_{t-1}$ (previous state). 1 represents completely saved, and 0 represents fully deleted. The calculation formula is:

$$F_t = Sigmoid(W_f [d_{t-1}, x_t] + b_f) \qquad (4)$$

Among them, $W_f$ is the weight matrix, $b_f$ is the bias term, and the output $F$ through this network is a number in the range (0, 1), indicating the probability of the previous cell state being forgotten, and 1 is "Completely reserved", 0 is "completely discarded".

The input gate in LSTM requires supplementing the newest memory from the current input after the circulating neural network "forgets" part of the previous state. This process needs to be fulfilled by the "input gate". The input gate consists of two parts. In the first part, a sigmoid layer named the "input threshold layer" decides which values we need to renew. The second part, a tanh layer, establishes a new candidate vector $\widetilde{C}_t$, which will be increased to this state. Its formula is as follows:

$$h_t = \sigma(W_n \bullet [d_{t-1}, X_t] + bn) \qquad (5)$$

$$\widetilde{C}_t = (W_M \bullet [d_{t-1}, X_t] + bm) \qquad (6)$$

$$C_t = F_t * C_{t-1} + h_t * \widetilde{C}_t \qquad (7)$$

Among them, $W_n$ represents the weight matrix, $b_n$ represents the bias item, $W_M$ represents the weight matrix for updating the state of the unit, $b_m$ represents the bias item for updating the state of the unit [46], and $C_t$ represents the state of the updated memory unit. In Formula (7), input gate $h_t$ and $\widetilde{C}_t$, do the dot product to decide whether to update the state of time-step memory unit; The forgetting gate $F_t$ takes the dot product with $C_{t-1}$ to decide whether to retain the original state of the time-step memory unit.

The output gate in LSTM is the output of the current moment that needs to be generated after calculating the new state, which is used to control how much the state of the memory unit in this layer is filtered. The output gate determines the output at that moment according to the latest state, the output at the last moment, and the current input. Its calculation formula is as follows:

$$d_t = O_t * \tanh(C_t) \tag{8}$$

$$O_t = \sigma(W_O[d_{t-1}, x_t] + b_O) \tag{9}$$

That is, first use the sigmoid activation function to obtain an $O_t$ with a value in the interval [0,1], and then multiply the memory cell state $C_t$ by the tanh activation function and then multiply it with $O_t$, which is the output of this layer. $d_t$ is not only related to the input $x_t$ under the time step $t$ and the activation value $d_{t-1}$ of the hidden layer in the previous time step, but also related to the memory unit state $C_t$ under the current time step.

## 4. Non-intrusive load decomposition based on hybrid deep learning

### 4.1. Hybrid CNN-LSTM model

Both CNN and LSTM are mainstream algorithms used in deep learning. CNN is good at extracting local features of data and acting on spatial abstraction and generalization. LSTM networks can expand time features and process data information with sequential features. Existing studies have shown that combining CNN and LSTM models is more stable than using CNN and LSTM models separately [47]. Therefore, this paper also associates the respective characteristics of the CNN network with the LSTM network and uses the parallel connection of the network to combine the two networks to obtain the CNN-LSTM network model of this article, which makes full use of the time and space feature expression capabilities of the two networks. The frame diagram of the CNN-LSTM parallel network is in Fig. 5.

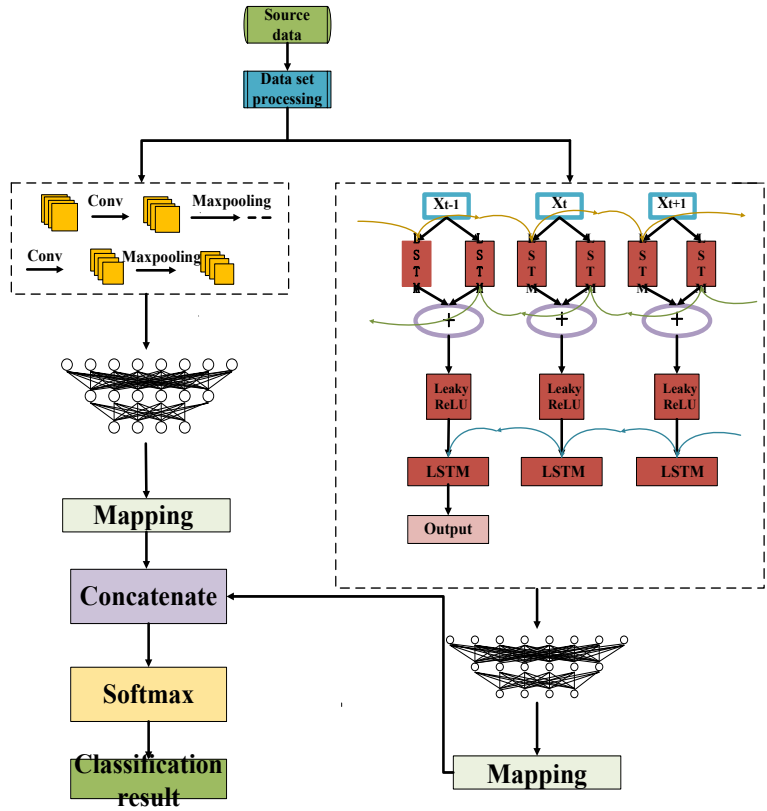

Fig. 5. Structure diagram of CNN-LSTM hybrid model

As shown in the figure, the data is preprocessed first, and the processed data is used to train the CNN and LSTM networks respectively. Next, the feature information extracted by CNN and the feature information extracted by LSTM are respectively processed into the same dimension through the mapping layer, and the outputs of CNN and LSTM are connected in parallel through concatenate, and finally classified by softmax.

During the experiment, some parameters have been defined for the model's architecture. These parameters are also called hyperparameters. For the purpose of finding the best accuracy of the hybrid model, it is first important to adjust these hyperparameters by trial and error.

According to the hybrid model method proposed, the parameters of the CNN and LSTM network are in Table 1. The number of training iterations is 200.

**Table 1** Parameter Settings

| Model category | The main parameters | The values |
|---|---|---|
| CNN | Size of the convolution kernel | 3×3 |
| | Number of convolution kernels in the first convolution layer | 64 |
| | Number of convolution kernels in the second convolution layer | 128 |
| | Step size | 1 |

|      | The activation function  | LeakyReLU |
|------|--------------------------|-----------|
|      | Optimization function    | Adam      |
|      | vector                   | 0.0001    |
| LSTM | Number of hidden neurons | 64        |
|      | Optimization function    | Adam      |
|      | vector                   | 0.0001    |

## 4.2. Evaluation indicators of the hybrid model

Taking into account that accuracy has some occasionality, statistical inditcators should be employed to comprehensively evaluate the performance of a classification model [48,49]. To this end, the evaluation indicators selected in this article are accuracy rate (ACC), F1-score, and MCC.

Accuracy (ACC) represents the proportion of the classification model that is correct for the entire sample. It is a very intuitive evaluation index. The formula is as follows:

$$ACC = \frac{TP+TN}{TP+TN+FP+FN} \quad (10)$$

Among them, $TP$ means that the predicted class is a positive class and is correctly judged as a positive class by the model, $FN$ means that the predicted class is a positive class but is wrongly judged as a negative class by the model, and $TN$ means that the predicted class is a negative class and is correctly judged by the model. The judgment is a negative class, $FP$ means that the predicted class is a negative class and is wrongly judged as a positive class by the model.

F1-Score is a calculation result that comprehensively premeditates the precision and recall of the model. The core idea is to improve Precision and Recall as much as possible but also hope that the difference between the two is as small as possible. The more natural the F1-score, the better the quality of the model, but the generalization capability of the model should be considered.

Precision refers to how many of all samples whose model is positive are real samples, and its formula is:

$$Precision = \frac{TP}{TP+FP} \quad (11)$$

Recall refers to how many positive examples in the sample are correctly predicted [48], and its formula is:

$$Recall = \frac{TP}{TP+FN} \quad (12)$$

From the above formulas of Precision and Recall, it can be seen that there may be a certain contradiction between the existence of the two. In order to take into account these two indicators, F1-Score is adopted in this study. The formula of F1-Score is as follows [50]:

$$F1 = 2 \bullet \frac{Precision \bullet Recall}{Precision + Recall} \quad (13)$$

MCC is very effective for the evaluation of unbalanced data sets. In essence, it is a correlation coefficient describing the predicted consequences and the true results. The closer the value is to 1, the more consistent the prediction is with the actual results. Its formula is as follows [12]:

$$MCC = \frac{TP \bullet TN - TP \bullet FN}{\sqrt{(TP+FP)} \bullet \sqrt{(TP+FN) \bullet (TN+FP) \bullet (TN+FN)}} \quad (14)$$

**4.3. Flow chart of CNN-LSTM hybrid model**

The flow chart of the hybrid model based on non-intrusive load decomposition proposed in this paper in Fig. 6. :

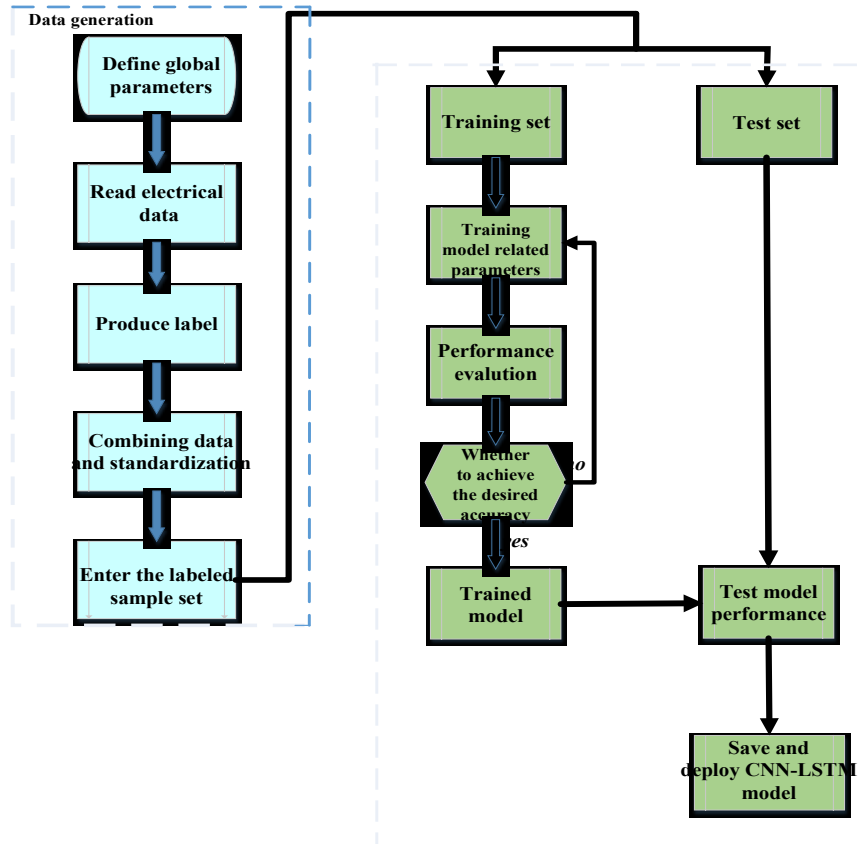

Fig. 6. Flowchart of the hybrid model

In the figure, the first is to read and preprocess the data set and use the labeled and processed sample set as the training set to train the parameters of the mixed model. When the accuracy of the model parameters used after training reaches the expected parameters, the obtained results are retained, and then the data of the test set is used for further testing of the parameter results. If the final result is close to the ideal value, the best hybrid model will be completed, and finally, the switch state of each electrical appliance will be judged.

## 5. Case study

The experimental environment in this article is MacBook Pro (13-inch, 2019), the processor is 2.4 GHz quad-core Intel Core i5, the memory is 8GB, and the graphics card is Intel Iris Plus Graphics 655 1536 MB.

### 5.1. Generation of data set and extraction of electrical appliances

This article uses the UK-DALE public data set released by the UK Energy Research Center in 2017 as an experiment on data. The data set is specifically designed to test the outcome of NILM [51]. A total of five records are recorded through current sensors, power recording, and other equipment. The load data of British households, including active power, current information, and electrical switch status. Uk-UKDALE includes data on electrical appliances in 5 households, and this article uses data from the first household. The data of the first household included 72 electrical appliances and 53 meters. The sampling information lasted from November 2012 to April 2017. Premeditating the bulky amount of data, this article extracts the useful functions of the first household electrical appliances after 2013.

This article selects nine electrical appliances, including refrigerators, kettles, televisions, home theater computers, washer dryers, microwave ovens, radios, fans, and computers, as the experimental research objects, and does not consider the decomposition when multiple electrical appliances exist at the same time. From each electrical appliance read before, select the electrical appliance to take out the same length, the length of the training set is 15000, and the data length of the test set is 5000. For label generation, the data of nine electrical appliances are integrated, and the useful work is added together to become the sum of useful work at each moment. The labels are also combined together, and each electrical appliance corresponds to a column. As shown in fig.7. below, at a certain moment, when the corresponding appliance is turned on, the position of the corresponding id is 1; and when it is turned off, it is 0.

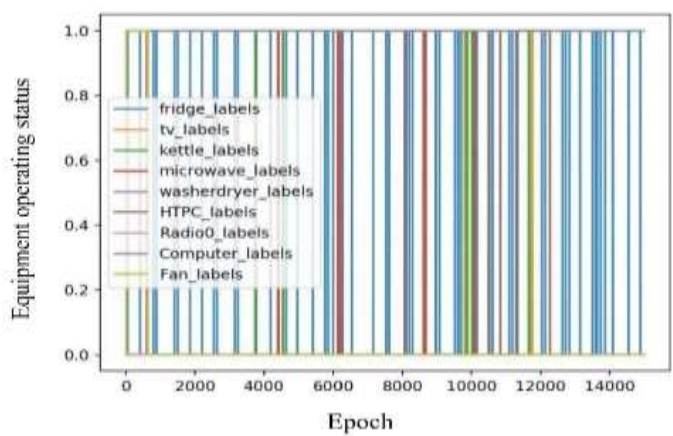

Fig. 7. Tag generation sequence diagram

### 5.2. Experimental results

The two line graphs below are the accuracy of the training set and test set respectively.

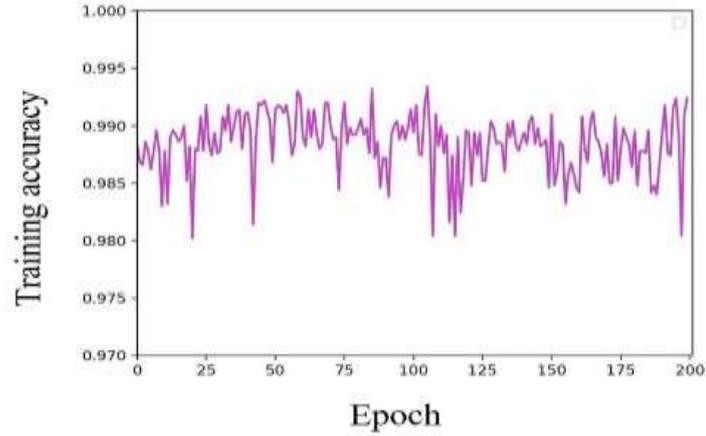

Fig. 8. The accuracy line diagram of the training set

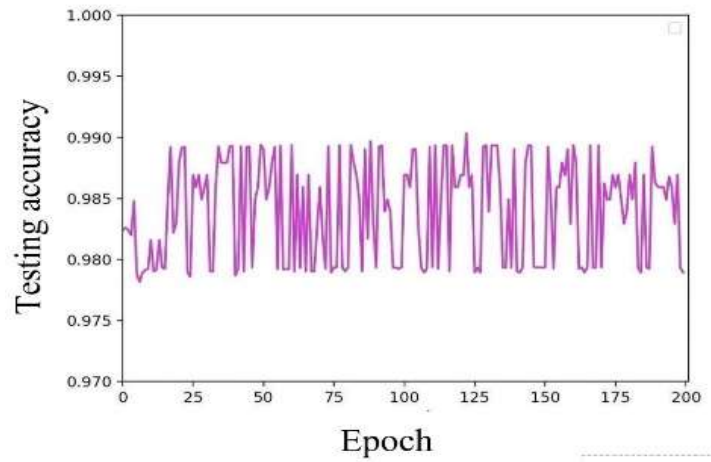

Fig. 9. The accuracy line diagram of the testing set

As can be derived from the figure, the accuracy of the training set fluctuates between 0.980 and 0.995, while that of the test set fluctuates between 0.978 and 0.990. The floating scope is small and the accuracy is high, which verifies the accuracy of this hybrid model.

The concrete data comparing the training and test accuracy of the hybrid model with the CNN, the LSTM network, and the Bidirectional GRU are made into the following table, as shown in Table 2:

Table 2 Comparison accuracy of each model

| Model | Training accuracy（%） | Test accuracy（%） |
| --- | --- | --- |
| CNN | 93.678 | 92.760 |
| LSTM | 95.826 | 93.673 |
| Bidirectional GRU | 92.982 | 90.854 |
| CNN-LSTM | 98.894 | 97.995 |

It can be derived from Table 2 that compared to a single CNN, LSTM network, and Bidirectional GRU network, the CNN-LSTM hybrid model has higher training accuracy and test accuracy.

### 5.3. Model performance evaluation

For the purpose of evaluating the performance of the mixed model, the confusion matrix of the model is first obtained in Table 3. It can be derived from the table that the false alarm rate of the hybrid model is 0.11%, and the false detection rate is 0.06%, so the proposed evaluation result is reliable.

**Table 3** Mixture model confusion matrix

| Confusion matrix | | The real value | |
|---|---|---|---|
| | | Positive | Negative |
| Predictive value | Positive | 1826 | 2 |
| | Negative | 2 | 3172 |

In addition, in order to evaluate the performance of the hybrid model more comprehensively and accurately, the hybrid model is compared with the other single two models from three different aspects: ACC, F1-score, and MCC in Table 4.

**Table 4** Performance comparison of each classification model

| Model | ACC（%） | F1-score | MCC |
|---|---|---|---|
| CNN | 97.86 | 0.994 | 0.986 |
| LSTM | 98.84 | 0.994 | 0.990 |
| CNN-LSTM | 99.96 | 0.998 | 0.998 |

It can be derived from Table 4 that, compared to other existing models [2], [17], [18], [52], the performance of the CNN-LSTM hybrid model in various aspects in a short time Both are better than existing models.

## 5.4. The sensitivity of the model with the change of computational parameters

In the proposed CNN-LSTM deep learning network, some system parameters significantly affect the evaluation performance, such as the learning rate or the number of the convolutional kernels. The sensitivity of these two main parameters is analyzed in this paper.

As an important super parameter in deep learning, the learning rate determines whether and when the objective function converges to the local minimum value. An appropriate learning rate can make the objective function converge to the local minimum value inappropriate time. Sensitivity analysis of the learning rate is shown in Table 5.

**Table 5** Sensitivity analysis of learning rate

| Learning rate | Training Accuracy | Test Accuracy |
|---|---|---|
| $1\times 10^{-3}$ | 96.692 | 95.599 |
| $1\times 10^{-4}$ | 98.868 | 97.695 |
| $1\times 10^{-5}$ | 97.236 | 96.568 |

|  |  | $1\times 10^{-6}$ | 96.025 | 95.230 |

As can be seen from Table 5, when the learning rate is $1\times 10^{-4}$, the training accuracy and test accuracy of the network reach the highest. Therefore, $1\times 10^{-4}$ is chosen as the learning rate in this paper.

The number of convolution kernels in CNN determines the depth of CNN, so choosing the different numbers of convolution kernels will affect the performance of the network. The sensitivity analysis of the number of convolution kernels is shown in table 6.

Table 6 Sensitivity analysis of the number of convolution kernels

| Number of convolution kernels | | Training Accuracy | Test Accuracy |
|---|---|---|---|
| The first layer | The second floor | | |
| 32 | 64 | 97.548 | 96.169 |
| 32 | 128 | 97.184 | 96.002 |
| 64 | 128 | 98.872 | 97.732 |
| 64 | 256 | 98.026 | 97.258 |

As can be seen from Table 6, the network achieves the best performance when the number of Conv kernels of two layers of CNN is 64 and 128 respectively.

**5.5. Calculation efficiency analysis**

For the purpose of better analyze the calculation efficiency of different models, we calculated the training time and test time of the hybrid network and compared it with the single network in Table 7.

Table 7 Time contrast of each classification model

| Model | Training time (s) | Test time (μs) |
|---|---|---|
| CNN | 1.92 | 23 |
| LSTM | 1.95 | 27 |
| CNN-LSTM | 2.08 | 1031 |

It can be derived from Table 7 that the training time of the CNN-LSTM hybrid network is almost the same as the training time of the single network. Although the test time is slightly longer, it is still within the acceptable range, and the hybrid network has a higher accuracy rate. Therefore, considering the time and efficiency, this paper chooses the CNN-LSTM hybrid network.

**5.6. Add noise to test the robustness of the model**

In electrical decomposition, noise is inevitable. For the purpose of better test the robustness of the mixed model, Gaussian white Noise with various Signal to Noise ratios (SNR) is increased in the data for the performance test.

The signal-to-noise ratio is designed to be 40dB, 30dB, and 20dB respectively. The robustness results are obtained under the hybrid model in Fig. 10.

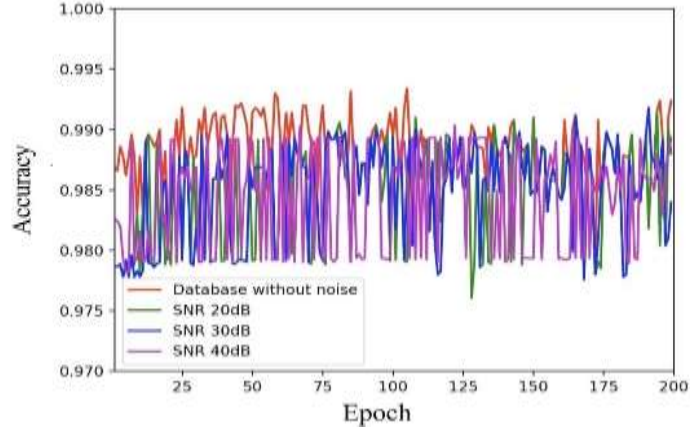

Fig. 10. Comparison of the robustness of the mixed model under three kinds of noises

It can be found from Figure 10 that compared with the original data without noise, the accuracy of the data with different kinds of noise is slightly reduced, but it still remains in a higher range.

As shown in Table 8 below, we also compared the three aspects of ACC, MCC, and F1-score for the various SNRs added.

Table 8 Performance comparison under different signal-to-noise ratios

| Signal to noise ratio | ACC (%) | F1-score | MCC |
| --- | --- | --- | --- |
| 20 dB | 99.56% | 0.9939 | 0.9900 |
| 30 dB | 99.62% | 0.9956 | 0.9912 |
| 40 dB | 99.86% | 0.9979 | 0.9962 |

It can be derived from Table 8 that under various signal-to-noise ratios, the hybrid model still maintains a high performance, which further proves the utility of the model.

## 6. Conclusion

By leveraging deep learning, this paper proposes a non-intrusive load decomposition method based on the CNN-LSTM hybrid model: 1) the proposed model manages to perform non-intrusive load decomposition tasks by making use of the spatial features of CNN and the temporal features of LSTM network. The advantage of this method is that it takes into account the characteristics of data feature extraction and is based on time series. The network structure is simple but powerful, and the recognition error under-sampling is reduced. In fact, most of this approach utilizes low-frequency data, so it is possible to implement the system using low-cost instrumentation to obtain aggregated load data. 2) In addition, we compare the proposed model with separate CNN and LSTM network models and use the three evaluation indicators of MCC, ACC, and F1-Score to comprehensively evaluate all models. At the same time, Gaussian noise with various signal-to-noise ratios is added to further test the robustness of the mixed model. 3) The simulations on the UK public data set UKDALE demonstrate that the presented approach yields an excellent decomposition result and outperforms the other alternatives.

For the shortcomings of the method proposed in this paper, the deep neural network often needs a large amount of data for training to avoid over-fitting. In this respect, we can expand the training data set by generating more synthetic data. Non-intrusive load decomposition makes it possible to know the appliance-level user behavior and measure the demand response potentials of flexible loads, which is beneficial to develop an advanced optimal scheduling method for residential microgrids or community integrated energy systems by considering flexible demand response [53-55]. In addition, it would be interesting to develop multi-objective NILM via advanced multi-objective heuristic algorithms [56,57] or design a NILM approach in a privacy-preserving manner by using federated learning [58]. Another interesting topic is to automatically determine the optimal hyperparameters of a deep learning-based NILM model by using automated machine learning [59,60].